\documentstyle[pre,aps]{revtex}        
\input epsf
\tighten
\begin{document}

\twocolumn[\hsize\textwidth\columnwidth\hsize\csname @twocolumnfalse\endcsname
\title{Correlation functions, free energies and magnetizations 
in the two-dimensional\\
random-field Ising model}
\author{
S. L. A. de Queiroz$^a$\footnote{Electronic address: sldq@if.ufrj.br}
and
R. B. Stinchcombe$^b$\footnote{Electronic address: stinch@thphys.ox.ac.uk}
 }
\address{
$^a$ Instituto de F\'\i sica, Universidade Federal do Rio de Janeiro,\\
Caixa Postal 68528, 21945-970  Rio de Janeiro RJ, Brazil\\
$^b$ Department of Physics, Theoretical Physics, University of Oxford,\\
 1 Keble Road, Oxford OX1 3NP, United Kingdom 
}
\date{\today}
\maketitle
\begin{abstract}
Transfer-matrix methods are used to calculate spin-spin
correlation functions ($G$), Helmholtz free energies ($f$) and
magnetizations ($m$) in the two-dimensional random-field
Ising model close to the zero-field bulk critical temperature $T_{c\ 0}$,
on long strips of width $L = 3 - 18$ sites, for binary field
distributions. Analysis of the probability distributions of $G$ for
varying spin-spin distances $R$ shows that
describing the decay of their  averaged values by effective correlation
lengths is a valid procedure only for not very large $R$. 
Connections between field-- and 
correlation function distributions at high temperatures are established,
yielding approximate analytical expressions for the
latter, which are used for computation of the corresponding
structure factor. It is shown that, for fixed $R/L$, the fractional widths of
correlation-function distributions saturate asymptotically with
$L^{-2.2}$. Considering an added uniform applied field $h$, a connection
between $f(h)$, $m(h)$, the Gibbs free energy $g(m)$ and the distribution
function for the uniform magnetization in zero uniform field, $P_0(m)$, is
derived and  first illustrated for pure systems, and then applied for
non-zero random field. 
From finite-size scaling and crossover arguments, coupled with numerical
data, it is found
that the width of $P_0(m)$ varies against
(non-vanishing, but small) random-field intensity $H_0$ as
$H_0^{-3/7}$. 
\end{abstract}

\twocolumn
\narrowtext
\vskip0.5pc]

\section{Introduction}
\label{intro}
The random-field Ising model (RFIM) has posed a number of challenges to
researchers since its introduction as an apparently purely theoretical
puzzle~\cite{imr75}. The later realization that it corresponds, give or
take a few (hopefully irrelevant) details, to the experimentally
realizable dilute Ising antiferromagnet in a uniform applied
field~\cite{fis79} brought new insights, and new questions as well;
among the latter, is the interpretation of experimental data in
a suitable theoretical framework. This has proved to be rather an
intricate subject, even down to basic aspects such as whether the
lower critical dimensionality for the problem was $d=2$ or 
$3$~\cite{yos82,bir83,bel83}. Though by now this particular issue has been
settled in
favour of $d=2$~\cite{imb84etc}, several important aspects 
(such as the scaling
behaviour near the destroyed phase transition in $d=2$,
which will be of interest here) still require further
elucidation~\cite{bel98}.
   
In the present paper we deal with the two-dimensional RFIM, where 
long-range order is destroyed, and a zero-temperature, zero-field
``anomalous'' critical point appears~\cite{aha83}.
The latter will not concern us directly, as we shall be working at
high temperatures, close to the pure-system ferro--paramagnetic
transition. We extend and complement our early work~\cite{rfcfd},
making use of transfer-matrix (TM) methods on long, finite-width strips of
a square lattice; we generate and analyze statistics of spin-spin
correlation functions and uniform magnetizations. Wherever feasible, we
attempt to draw connections between our numerical results and
experimentally observable quantities. 
In what follows, we begin by briefly reviewing selected aspects of the
numerical techniques used, and how they relate to the
physical problem under study. We then recall the connection between
structure factors and averaged correlations in random systems, and discuss
the extraction of 
effective correlation lengths from our numerical data
for correlation-function statistics. Next we exploit
the connection between field-- and 
correlation function distributions at high temperatures, in an
attempt to derive approximate analytical expressions for the latter; such
formulae are used in turn, in order to compute the corresponding structure
factor. A short section is dedicated to a reanalysis of the
asymptotic behaviour of the widths of correlation-function distributions,
first presented in Ref.~\onlinecite{rfcfd}, and now complemented by
additional data.
In the next section, an additional uniform applied field is considered:
free energies and uniform magnetizations are calculated on
strips of both pure and RFIM systems. These quantities are used 
to calculate the corresponding Gibbs free energy which, in turn, gives
the distribution function for the uniform magnetization in zero uniform
field. Numerical data are then analyzed via
finite-size scaling and crossover arguments. 
A final section summarizes our work.

\section{Numerical methods and $d=2$ RFIM}
\label{secII}
We consider  strips of a square
lattice of ferromagnetic Ising spins with nearest-neighbour interaction
$J=1$, of width $3 \leq L \leq 18$ sites with periodic boundary conditions
across.
The random-field values $h_i$  are drawn for each site $i$ from the
binary distribution:
\begin{equation} 
p(h_i) = {1 \over 2}[\, \delta (h_i - H_0) + \delta (h_i + H_0)\, ]\ \  .
\label{eq:pd}
\end{equation}
TM methods are used, on long strips of typical length $L_x=10^6$ 
columns, as described at length in Ref.~\onlinecite{rfcfd} and
references therein, to generate representative samples of the
quenched random fields. Along the strip, we calculate correlation
functions (as explained
in the next paragraph), as well as free energies and magnetizations
(details in Section~\ref{mag}).

Here we calculate the disconnected spin-spin correlation function 
$G(R) \equiv \langle \sigma_0^1 \sigma_R^1 \rangle$, between spins on the
same row (say, row 1), and $R$ columns apart.  Related quantities, such as 
correlation lengths,  are defined with
connected correlations, $\langle \sigma_0 \sigma_R \rangle-
\langle \sigma \rangle^2 $, in mind; however, for the
quasi-one-dimensional Ising systems under consideration (either pure or
random) one
is always in the paramagnetic phase, so the distinction between connected
and disconnected correlations is unimportant.
In Ref.~\onlinecite{rfcfd} we 
explained why the ranges of spin-spin distance, temperature
and random-field intensity of most interest for investigation by TM
methods are, respectively, $R/L \simeq 1$, $0 <T \lesssim T_{c\
0} = 2.269 \cdots$ [we take $k_B \equiv 1$], $H_0 \lesssim 0.5 $.
Here we restrict ourselves to high 
$T \gtrsim 2.0$, and rather low fields, $H_0 \lesssim 0.1 - 0.15$ .
We use a linear binning for the histograms of occurrence of $G(R)$;
usually the whole $[-1,1]$ interval of variation  of $G(R)$ is
divided into $10^3$ bins.

Since we shall be dealing with probability distributions, a word is
in order about multifractality. Though multifractal
behaviour has been found {\it at} the critical point of random-bond Potts
systems~\cite{oy99,cha00}, the available evidence strongly suggests that,
off bulk criticality, correlation functions behave normally~\cite{oy99}.
Thus, in the present
case we expect that analysis of different moments of the probability
distribution of $G$ will yield essentially the same results. 

\section{Correlation decay}
 
The properties of correlation functions are usually incorporated
into associated  correlation lengths, whose basic
definition is as (minus) the inverse slope of semi-logarithmic plots of
correlation functions against distance. In this view, one assumes both 
that exponential decay can be well-defined at essentially all distances, 
and that a single length is enough to characterize such behaviour.
In cases as the present, quenched randomness
implies that configurational averages must be taken, and one must be
careful in deciding what quantities are to be thus promediated. 
Recall that, e.g. in neutron scattering experiments, the intensity of
the magnetic critical scattering is proportional to the
average (over the crystal) of the scattering function $S({\vec q}\,)$, 
which is the Fourier
transform of the correlation function for wave-vector transfer 
${\vec q}\,$~\cite{col89,sla00}. With $G_R \equiv G(R)$, and wave vector 
$q$ in the row direction, $S$ becomes
\begin{equation}
\large[ S(q)\large] = 
\large[ \int dR\, e^{iqR}G_R \large] =
 \int dR\, e^{iqR}\,\langle G_R \rangle
\label{eq:s(q)}
\end{equation}
where $\large[ \cdots\large]$ stands for configurational average, and
\begin{equation}
\langle G_R \rangle = \int dG_R\,P(G_R)\,G_R \ ,
\label{eq:avg}
\end{equation}
where $P(G_R)$ is the probability distribution for $G_R$. 
The last equality in Eq~(\ref{eq:s(q)}) depends only on the assumption
that $P(G_R)$ is position-independent along the crystal.

The simplest assumption for $P(G_R)$ that incorporates both disorder
and exponential decay given by a single length $\xi$ for all distances is
a Gaussian distribution:
\begin{equation}
P(G_R) = {1 \over \sqrt{2\pi}\,\Delta(R)}\,e^{-y^2/2\Delta^2(R)},\ \ y
\equiv G_R - e^{-R/\xi}
\label{eq:pgauss}
\end{equation}
where distance--dependent widths $\Delta(R)$ allow for, e.g.,
(disorder-induced) larger uncertainties for larger spin-spin
separations. However, using Eq.~(\ref{eq:pgauss}) in  
Eqs.~(\ref{eq:s(q)})--(\ref{eq:avg}) one
obtains a width-independent Lorentzian form for the average structure
factor:
\begin{equation}
\large[ S(q)\large] =  {1 \over q^2 + {1 \over \xi^2}}\ .
\label{eq:sgauss}
\end{equation}
This coincides with the standard mean-field result for the disordered
phase, and is deemed unsatisfactory upon comparison
with experimental data~\cite{bel98,sla00}.
\begin{figure}
\epsfxsize=8,4cm
\begin{center}
\leavevmode
\epsffile{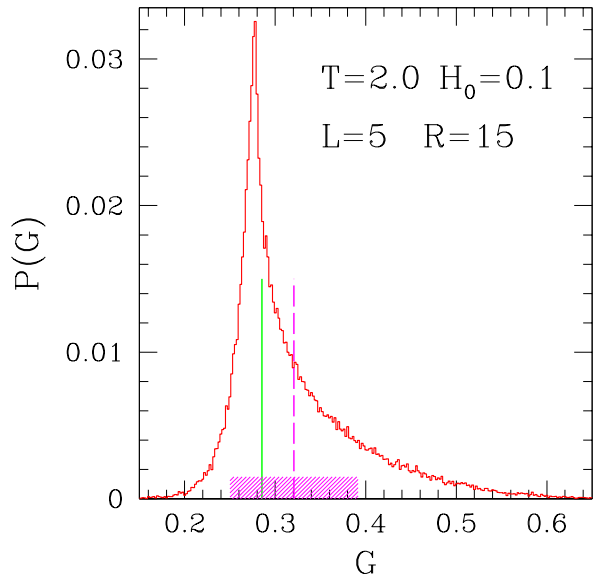}
\caption{
Normalized histogram $P(G)$ of occurrence of $G$. Strip length $L_x=10^6$
columns, binwidth $2 \times 10^{-3}$. Vertical bars located respectively 
at: $G_0$ (full line), $\langle G \rangle$ (dashed). Shaded region on
horizontal axis from  $\langle G \rangle-{\tilde W}$ to  $\langle G
\rangle+{\tilde W}$.}
\label{fig:hight}
\end{center}
\end{figure}
We now exhibit our numerical results, and compare their implications
to those of Eqs.~(\ref{eq:pgauss}) and~(\ref{eq:sgauss}). For high
temperatures and
low random-field intensities, as specified above, we recall (see also
Ref.~\onlinecite{rfcfd})
the following main features  found for the probability distribution
$P(G)$: (i) a clearly-identifiable, cusp-like peak,
at some $G_m$ {\em below} the zero-field value $G_0 \equiv G(H_0=0)$; (ii)
a short tail below the peak and a long one above it, such  that
(iii) all moments of order $\geq 0$ of the distribution are {\em above}
$G_0$. In Figure~\ref{fig:hight}, where the 
first moment $\langle G \rangle$ is shown, one has $G_m=0.278$, $G_0=
0.2853$, $\langle G \rangle=0.321$; the RMS width 
${\tilde W} \equiv \langle (G-\langle G \rangle)^2 \rangle^{1/2}=0.071$.

\begin{figure}
\epsfxsize=8,4cm
\begin{center}
\leavevmode
\epsffile{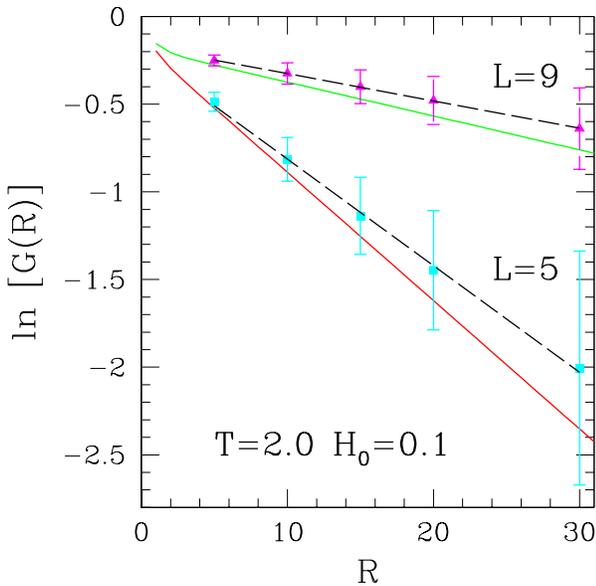}
\caption{
Correlation decay along strips of widths $L=5$ and $9$. Full lines: 
$H_0=0$. Points: $\langle G \rangle$ for $H_0=0.1$. Dashed lines:
unweighted least-squares fits of $H_0=0.1$ data. Vertical
bars give RMS widths ${\tilde W}$ of corresponding distributions.   
}
\label{fig:corrdec}
\end{center}
\end{figure}

Therefore, the features depicted in Fig.~\ref{fig:hight},
especially the asymmetric cusp, are
at variance with the form Eq.~(\ref{eq:pgauss}).
We now investigate what effects are carried over to the associated
correlation lengths.
We do so by mimicking the procedure outlined in Eqs.~(\ref{eq:s(q)})
and~(\ref{eq:avg}) above: first we average over randomness
for a given spin-spin separation, and then study the variation of
the averaged quantities over distance. 
The results are in
Fig.~\ref{fig:corrdec}, where our numerical data for
$\langle G \rangle$ are plotted against varying $R$. $H_0=0$ data
are also shown for comparison.  
One sees from the respective slopes that, taking into account data for $R
\gtrsim L$, the correlation length for $H_0
\neq 0$ would seem to be systematically
{\it larger} than in zero field. This reflects the domain structure into
which the system breaks down: at short distances the conditional
probability for a spin to belong to the same domain as the one at the
origin is {\it larger} than for $H_0=0$.
 
For longer distances, correlation functions start to show severe 
disorder-induced fluctuations, related to the crossing of domain walls.
Contrary to the zero-field case, where temperature-induced domain walls 
are present but the respective sign changes in correlation functions
average out to give an exponential fall off, here 
the domain wall configurations are essentially determined by the
(quenched) 
accumulated random-field fluctuations. At large $R$, such fluctuations
play a very sensitive role even for very low
random field intensities .
One anticipates problems with defining
correlation lengths from the corresponding data. 
The vertical bars in Fig.~\ref{fig:corrdec} show that, 
for fixed strip width $L$,
the width ${\tilde W}$ of the distribution indeed grows apparently
unbounded 
for increasing $R$; though this is related to the crossing of domain walls
just mentioned, it is also, and predominantly,
an intrinsic feature of the quasi-one-dimensional systems used
here. Thus, inferring two-dimensional behaviour from such trends may be
risky. However, we now argue that in $d=2$ one does actually run into
problems
for large $R$, exactly as inferred above; only, the underlying reasoning
is subtler. 
   
In fact, see Ref.~\onlinecite{rfcfd} and Section~\ref{widths} below,
a different analysis of correlation functions, at fixed $R/L$, strongly
suggests that the 
{\it relative} widths $W \equiv {\tilde W}/\langle G \rangle$ grow as 
$R, L \to \infty$ in $d=2$, 
approaching a finite limiting value $C\,H_0^\kappa,\ C \simeq 2, \kappa
\simeq 0.5$.
This means that, when one considers the dispersion of
$\ln \langle G \rangle$,  the signal-to-noise ratio
becomes of order one for large $R, L$, and it is this latter
fact that, in $d=2$, must compromise
attempts to extract correlation lengths in such range.

The effect of the above on fits of neutron-scattering data to lineshapes
is that, since the latter rely on the idea that correlation lengths are
always reliable quantities, they may be off the actual picture in
the small-wavevector region. 

We now attempt to derive approximate analytical expressions for $P(G_R)$;
our ultimate goal is to predict a form for $\large[ S(q)\large]$ from
Eqs.~(\ref{eq:s(q)})--~(\ref{eq:avg}).

\section{Distribution of $G$ from field distribution (at high T)}
\label{secIII} 

In this section we use simple scaling ideas to establish a quantitative
connection between the underlying distribution of
accumulated fields and that of the correlation functions themselves.

We begin by considering a one-dimensional system with sites denoted
by $i=0,1, \ldots$, uniform
nearest-neighbour interactions $K$ and site-dependent random fields
$h_i$ (both in units of $T$). For any given specific realisation of the
fields an 
exact decimation scheme with length
scaling factor $b=2$ can be applied, eliminating all odd-numbered 
sites. The renormalised fields at, and coupling between, e. g.
spins $0$ and $2$, are given by:
\begin{eqnarray}
h_0^{\prime} - h_2^{\prime} &=& h_0 - h_2 \nonumber \\
h_0^{\prime} + h_2^{\prime} &=& h_0 + h_2 +\frac{1}{2} \ln\left[
\frac{\cosh 2(K+h_1)}{\cosh 2(K-h_1)}\right]\nonumber\\
4 K^{\prime} &=& \ln\left[\frac{\cosh 2(K+h_1)\cosh 2(K-h_1)}
{\cosh^2 2h_1}\right]\ \ \ .
\label{eq:dec}
\end{eqnarray}
Iterating this procedure $n$ times, one obtains a single renormalised
bond $\tilde K$ connecting sites $0$ and $R$ ($R = 2^n$), at
which
the respective rescaled fields are $\tilde h_0$, $\tilde h_R$. The
correlation function $G(R) \equiv \langle \sigma_0 \sigma_R \rangle$ is
therefore:
\begin{equation}
G(R)=
\frac{e^{2\tilde K}\cosh(\tilde h_0 + \tilde h_R) - \cosh(\tilde h_0 -
\tilde h_R)}{e^{2\tilde K}\cosh(\tilde h_0 + \tilde h_R) + \cosh(\tilde
h_0 - \tilde h_R)}
\label{eq:g(r)}
\end{equation}
For low fields $H_0 \ll 1$, one uses 
$\cosh(\tilde h_0 - \tilde h_R) /\cosh(\tilde h_0 + \tilde h_R) \simeq
\exp(-2\tilde h_0  \tilde h_R)$ to get
\begin{equation}
G(R) \simeq \tanh(\tilde K + \tilde h_0  \tilde h_R)\ .
\label{eq:approx}
\end{equation}
Then, provided also $H_0 \ll K$, the distribution of $G(R)$ is given by
that of $X \equiv
\tilde h_0  \tilde h_R$, since (to lowest order in $H_0$) 
$K^\prime= {1\over 2}\ln\cosh 2K$ is field-independent. One has:
\begin{equation}
P(X)=
\int d\tilde h_0\,\tilde P (\tilde h_0)\,\int d\tilde h_R\,\tilde
P(\tilde h_R)\, \delta(X-\tilde h_0  \tilde h_R)\ .
\label{eq:xdist1}
\end{equation}
At low $H_0$, the scaling equations~(\ref{eq:dec}) give $h_0^{\prime} \sim
h_0 + h_1\,\tanh 2K$. Repeated applications of this transformation give
$\tilde h_0 \sim \sum_{i=1}^R h_i$ if $R \ll \xi$, where $\xi \sim \ln
(\tanh K)^{-1}$ is the correlation length at low $H_0$. Then $\tilde h_0$
(and similarly $\tilde h_R$) is the sum of ${\cal N}$ independent
variables (${\cal N}=R$), so the individual distributions of $\tilde h_0$,
$\tilde h_R$ become (at large $R$, $\xi$) Gaussians of width $\Delta
\equiv H_0{\sqrt{\cal N}}$:
\begin{equation} 
\tilde P (\tilde h_{0,R}) \propto
\exp\left(-(\tilde h_{0,R}/\Delta)^2\right)\ .
\label{eq:xdist1a}
\end{equation}
For $R \gtrsim \xi$, the same form applies, but because the field
accumulation under scaling is cut off by the decreasing $\tanh K$,
the relation for $\Delta$ involves ${\cal N} \sim \xi$. So, in
general, ${\cal N} \sim {\rm min}\,(R,\xi)$.

Making $\tilde h_0 = s\,\cos\theta$, $\tilde
h_R = s\,\sin\theta$,
\begin{equation}
P(X) \propto \int_0^\infty ds\,s\,\int_0^{2\pi}d\theta\,
e^{-s^2/\Delta^2}\,
\delta(X- {s^2 \over 2}\sin 2\theta)
\label{eq:xdist2}
\end{equation}
with the final result
\begin{equation}
P(X)= {a \over \Delta^2}\, e^{-y}\,\ln \left( 1 +{1 \over
y}\right)\ , \qquad y \equiv {2 |X| \over \Delta^2}\ 
\label{eq:xdist3}
\end{equation}
where $a$ is an overall normalization constant. Strictly
speaking, Eq.~(\ref{eq:xdist3}) is the {\em asymptotic} reduction of
Eq.~(\ref{eq:xdist2}), valid for the regimes $y \ll 1$ (the relevant
one for our purposes, as shown below) and $y \gg 1$.\par\noindent
Transforming back to $P(G)$, one sees that the value\ $G_m$ for which
$P(G)$ is maximum must correspond to $X=0$, which maximizes $P(X)$.
Thus, from Eq.~(\ref{eq:approx}),
\begin{equation}
y= {2 \over \Delta^2}\left|
\tanh^{-1}G_m -\tanh^{-1}G \right|
\label{eq:y/g}
\end{equation}
For $G$ close to $G_m$, linearization gives
\begin{equation}
P(G) \sim {\exp\left[ -{1 \over \tilde \Delta^2}\left|
G_m-G \right|\,\right] \over 2 \Delta^2
\left[1-G^2\right]}\ln\left[1+{\tilde \Delta^2 \over \left|
G_m -G \right|}\right]\ ,
\label{eq:gdist2}
\end{equation}
with $\tilde \Delta^2 = {1 \over 2} \Delta^2(1-G_m^2)$.

\begin{figure}
\epsfxsize=8,4cm
\begin{center}
\leavevmode
\epsffile{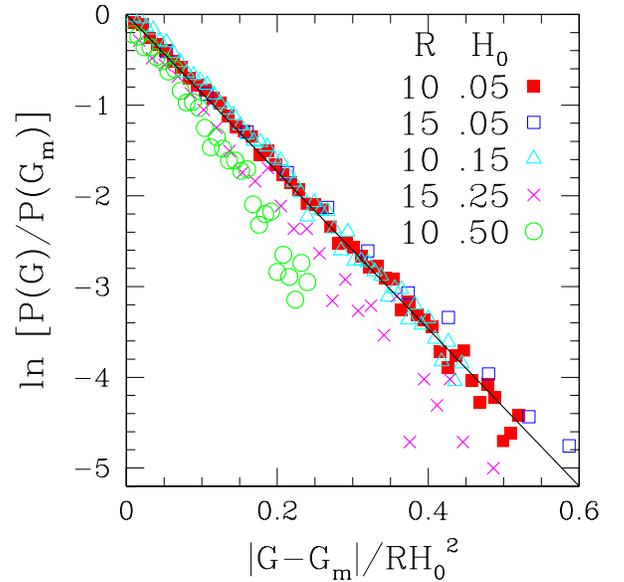}
\caption{Semilogarithmic plots of distribution functions below $G_m$,
the value at which $P(G)$ peaks
(see text). $L=5$, $T=2.0$. Straight line is guide to the eye.
}
\label{fig:slcoll}
\end{center}
\end{figure}

The main feature exhibited by this form is a locally
symmetric cusp, with infinite
slope on either side, at $G=G_m$. This is expected to carry
over to more general contexts, provided that $H_0 \ll 1$. Indeed
we have checked that a similar description (applying approximate
Migdal-Kadanoff scaling calculations), with the prediction of a cusp,
also applies on strips and in two dimensions (see Eqs.~(\ref{eq:mk1}),
(\ref{eq:mk2}) below, and related discussion).
A  quantitative test of Eq.~(\ref{eq:gdist2}) is
shown in Figure~\ref{fig:slcoll}, where only data for $G \leq G_m$ are
displayed (we shall deal with $G > G_m$ immediately afterwards). 
The conditions are such that $R \lesssim \xi$, $G_m^2 \ll 1$ (see 
Fig.~\ref{fig:corrdec}), so the
above low-field theory gives $\tilde \Delta \sim H_0\sqrt{R}$.
One sees that exponential decay against $|G-G_m|/RH_0^2$ is indeed
the dominant behaviour, provided that $H_0 \lesssim 0.15$; already for
$H_0=0.25$, small departures show, which become more prominent for
$H_0=0.5$.

As regards cusp asymmetry, not predicted by Eq.~(\ref{eq:gdist2}), we have
found that although data for $G>G_m$
still fall exponentially for small $H_0$, they do not collapse when
plotted  against $|G-G_m|/RH_0^2$. This is because the mutual
reinforcement, between
ferromagnetic spin-spin interactions and accumulated field fluctuations
(responsible for the long forward tail~\cite{rfcfd}), is left out by the
approximation above Eq.~(\ref{eq:xdist1}), namely that $K^\prime$ is 
$H_0$-independent. 

Before calculating $\large[ S(q)\large]$ from Eq.~(\ref{eq:gdist2}),
we recall that Eqs.~(\ref{eq:s(q)})--(\ref{eq:avg}) are
normally required for
bulk systems, thus one must work out an approximate scheme
to go from the $d=1$ regime of  Eqs.~(\ref{eq:dec})--(\ref{eq:gdist2}) to
$d=2$. We have
done so via a Midgal-Kadanoff rescaling transformation at 
$T \sim T_{c\ 0}$.
As a consequence of the similarity of the
corresponding recursion relations, to those for one dimension,
one ends up, after $m$ scalings such
that $2^m=R$, with a result very similar to  Eq.~(\ref{eq:approx}):  
\begin{equation}
G(R) = \tanh(\tilde K + \tilde h_0  \tilde h_R)\ ,
\label{eq:mk1}
\end{equation}
where again one assumes low fields,  $\tilde h_0$, $\tilde h_R$ 
For large $R$ these have Gaussian distributions of width $\Delta_R$
determined by the eigenvalue $\lambda$ of the low-field scaling
transformation of $H_0$. For $T$ near $T_{c\ 0}$, $\Delta_R \propto
R^\mu\,H_0$ where $\mu =\ln \lambda / \ln b$ ($b=2$). Further, 
\begin{equation}
\tilde K  \sim K_c -{R \over \xi}\ .
\label{eq:mk2}
\end{equation}
Since Eqs.~(\ref{eq:xdist1})--(\ref{eq:xdist3}) still apply, provided
$\Delta$ is replaced by $\Delta_R$,
one gets
the dominant contribution to the scattering function as:
\begin{equation}
\large[ S(q)\large] \propto {\rm Re} \int dR\,e^{iqR}\,e^{-R/\xi}
\left(1+ C\,\Delta_R^2\right)\ \ ,
\label{eq:s(q)pred}
\end{equation}
where $C$ is a constant of order unity. 
This can be transformed into:
\begin{equation}
\large[ S(q)\large] \propto {1/\xi \over {1 \over \xi}^2 +q^2} +
C\,H_0^2 (2\mu)!\ {\rm Re}\left( {1 \over \xi} -iq\right)^{-(2\mu+1)}\ .
\label{eq:s(q)pred2}
\end{equation}
If one assumes the form $\Delta =H_0 \sqrt{R}$,
given for $R \leq \xi$ in one dimension (see above and below
Eq.~(\ref{eq:xdist1a})), and also by the Migdal-Kadanoff scheme in $d=2$,
then $\mu =1/2$ and Eq.~(\ref{eq:s(q)pred2})
predicts the lineshape to be Lorentzian plus Lorentzian-squared, the
mean-field form found when the disconnected contribution is taken into
account~\cite{bir83,bel98,col89,sla00,gla86}. 
On the other hand, if ones goes by the
saturation behaviour predicted in Ref.~\cite{rfcfd} and
Section~\ref{widths}
below, and by the scaling approaches if $R \gtrsim \xi$, then
the result is $\zeta=0$, corresponding to a single Lorentzian in
Eq.~(\ref{eq:s(q)pred2}). 

Though either of these final predictions is certainly open
to challenge, in view of the number and severity of approximations 
involved in the course of their derivation, it is expected that the
procedure
described above will serve as a rough guide to attempts at connecting
basic microscopic features (such as fluctuations of accumulated fields)
to observable quantities, e.g. scattering functions.  

\section{Widths of $G$-distribution}
\label{widths}

In Ref.~\onlinecite{rfcfd} we studied the variation of the 
RMS relative width $W$ of the
probability distribution of correlation functions, against field
intensity and strip width, for fixed $R/L$, high temperatures and small
$H_0$. We proposed the scaling form
\begin{equation}
W = H_0^{\kappa}\, f(L\, H_0^u)\ \ \ ,
\label{eq:wscale} 
\end{equation}
and showed that, for $R/L=1$, $T=T_{c\ 0}$,
good data collapse of $y \equiv \ln \left[W\, H_0^{-\kappa}\right]$
against $x \equiv L\, H_0^u$ can indeed be obtained with $\kappa \simeq
0.43 - 0.50$ and $u \simeq 0.8$.
We used $L\leq 15$ and scanned $0 <x \lesssim 2.8$; keeping
$\kappa=0.45$ and $u=0.8$, we
found for $x>1$ a satisfactory fit given by $y=-0.3-5.3\,\exp(-1.57x)$,
which would imply an exponential saturation of the scaled width $W\,
h_0^{-\kappa}$ as $x \to \infty$, with a limiting value $\exp(-0.3) =
0.83$.
 
\begin{figure}
\epsfxsize=8,4cm
\begin{center}
\leavevmode
\epsffile{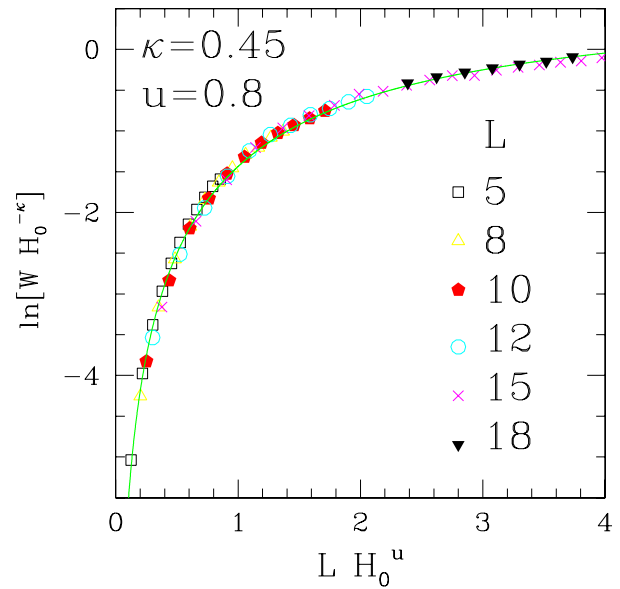}
\caption{Semi-logarithmic scaling plot of RMS relative widths, 
$W\, H_0^{\kappa}$
against $L\, H_0^u$.
Curve is fitting spline, given by Eq.~(\protect{\ref{eq:newfit}}).
}
\label{fig:wvslh}
\end{center}
\end{figure}

In Figure~\ref{fig:wvslh}, we display again the data of
Ref.~\onlinecite{rfcfd}, plus additional data for $L=15$ and $18$, 
which enabled us to explore
larger values of $x$  ($x \lesssim 4.0$) while still
keeping to relatively low $H_0$. We then reanalyzed our full set of data, 
with the results that (i) we managed an excellent 
fit to the whole interval $0<x<4$  by a single expression  
\begin{equation}
y=-2.2\,\left\{\ln\left({1.0 \over x}+0.6\right)\right\} -0.4025\ \ ,
\label{eq:newfit}
\end{equation}
(in which $y$ and $x$ involve the same values of the exponents
$\kappa$, $u$ as earlier)
and that (ii) this new fit, while still predicting saturation for
 $x \gg 1$, implies that in the approach to
two dimensions $x \gg 1$, convergence of $W\, h_0^{-\kappa}$
is power-law--like, giving a limiting scaled width 
$W\, h_0^{-\kappa} = 0.6^{-2.2}\,\exp(-0.4025) = 2.06$.

\section{Magnetizations}
\label{mag}
In this section we examine the scaling properties of the uniform
magnetization on strips of the $d=2$ RFIM. For convenience we shall always
keep $T =T_{c\ 0}$. 

We first outline our method, which involves a generalised Legendre
transformation.
Consider the Hamiltonian
\begin{equation}
{\cal H} = {\cal H}_0\left(\{\sigma\}\right) - h\sum_i\sigma_i
\label{eq:hmag}
\end{equation}  
where $\sigma_i$ are Ising spins, and ${\cal H}_0$ includes all
interactions except that of the spins with the uniform field $h$.
One has, for the corresponding partition function $Z(h)$:
\begin{equation}
{Z(h) \over Z(0)} = \sum_M P_0(M)\,e^{\beta Mh}\ ,\ \ \ \beta = {1
\over T}
\label{eq:p0m}
\end{equation}
where $P_0(M)$ is the probability of occurrence of the value $M$ for the
magnetization, {\em in zero uniform field}. Assuming a system with 
$N \gg 1$ spins,  with $f(h) \equiv$
negative free energy per site in units of $T$, and $m \equiv \beta
M/N$,
\begin{equation}
e^{N \left(f(h)-f(0)\right)} = N\int dm\,P_0(m)\,e^{N\,m\,h}\ .
\label{eq:fhN}
\end{equation}
In order for extensivity to be satisfied, one must have $N\,P_0(m) = \exp
N\,g(m)$, where $g(m)$ is intensive and determined by
\begin{equation}
e^{N \left(f(h)-f(0)\right)} = \int dm\,e^{N\left(g(m)+\,m\,h\right)}\ .   
\label{eq:fhN2}
\end{equation}
Assuming the usual sharp-peaked distribution around a
thermodynamically averaged value ${\overline m}$, one sees that
\begin{equation}
f(h)-f(0)= g({\overline m})+{\overline m}\,h +{\cal O}\left({\ln N \over
N}\right)
\label{eq:fhN3}
\end{equation}
with $(dg/dm)_{\overline m}=-h$. 
That is, $g$ is the standard Gibbs free energy per site. Substituting 
back in Eq.~(\ref{eq:fhN}), one gets:
\begin{equation}
P_0(M)=\exp N g(m)\ ,
\label{eq:fhN4}
\end{equation}
where terms of ${\cal O}(\ln N /N)$ have again been neglected.

Eq.~(\ref{eq:fhN4}), with $g(m)$ given through Eq.~(\ref{eq:fhN3}), is the
natural starting point to study magnetization
distributions by TM methods. Indeed, though one can get the
thermodynamically averaged exact values of all {\em moments} of the
distribution via TM~\cite{bd85,kb93}, the distribution itself is not
given directly.
This contrasts with Monte-Carlo methods, which inherently incorporate
readily-observable fluctuations around equilibrium, and have been widely
used to study magnetization distributions at criticality, both in
hypercubic geometries~\cite{magdist1} 
and on planar lattices with various aspect ratios~\cite{magdist2}.

Recall that, on strips of width $L$ and
length $L_x$, $L_x \gg L$ such as is the case here, the aspect ratio is
essentially infinite, therefore $P_0(M)$ will be Gaussian, at least for
pure systems~\cite{kb93,magdist2}. 
Our purpose (as shown below)
is to compare pure-- and RFIM-- results and explain their
mutual differences, by using
general theory of RF
systems~\cite{fis79,aha83,aha78,fer83,bel85}
coupled with finite-size scaling (FSS)~\cite{bar83}. 

\subsection{Pure Ising systems}
\label{pure}

We start by illustrating the properties of $g(m)$ for pure Ising spins.
One calculates $f(h)$, $f(0)$, $m(h)$ in Eq.~(\ref{eq:fhN3}) by standard
numerical methods~\cite{bn82}: the first two by isolating the largest
eigenvalue $\Lambda_0$ of the TM and using $f=L^{-1}\,\ln \Lambda_0$
(which is tantamount to assuming $L_x \to \infty$; more on this below), 
the third by calculating derivatives of $f$
relative to $h$. The latter is done here by perturbation
theory~\cite{bd85,kb93,sd85,dss87},
both for better numerical accuracy and because an adapted procedure
proves convenient when dealing with the RF case, where samples over
disorder must be accumulated.
\begin{figure}
\epsfxsize=8,4cm
\begin{center}
\leavevmode
\epsffile{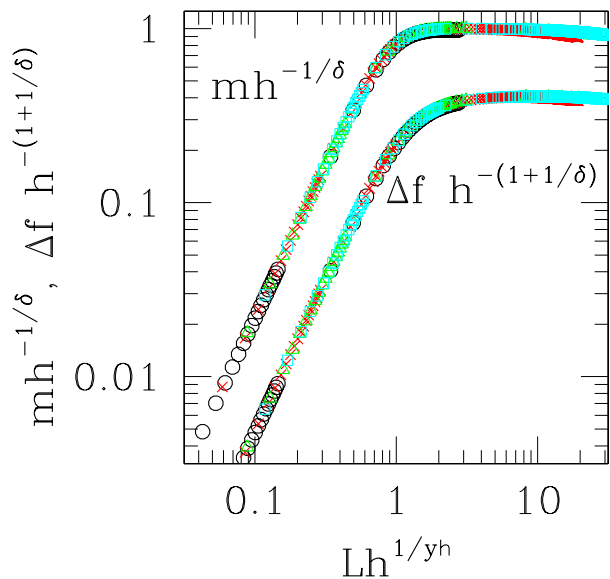}
\caption{
Scaling plots of magnetization and excess free energy for pure Ising
systems at criticality. Strip widths $L=4$
(circles), $8$ (crosses), $12$ (triangles), $16$ (squares).
Normalized magnetizations ($=T_c\, \partial f/\partial h$)
are used to avoid superposition of plots.
}
\label{fig:fmpure}
\end{center}
\end{figure}
At $t\equiv (T-T_c)/T_c=0$, FSS~\cite{bar83} gives for the excess free
energy: $\Delta f(h,L)\equiv f(t=0,h,L)-f(0,0,L)=
h^{1+1/\delta}\,F(L\,h^{1/y_h})$, with $\delta=15$,
$y_h=15/8$. In Fig.~\ref{fig:fmpure} we show scaling plots of 
$\Delta f(h)\,h^{-(1+1/\delta)}$ and $m\,h^{-1/\delta}$
against $L\,h^{1/y_h}$. For low fields ($h \lesssim L^{-y_h}$), the slopes
of both logarithmic
plots are given by the (finite-size) initial susceptibility exponent
$\gamma/\nu=7/4$,
as a consequence of the scaling relation $y_h(1-1/\delta)=\gamma/\nu$.

For $t$ non-zero, but 
still for low fields, one generally expects $\Delta f(h)= a(t,L)\,h^\mu$
whence $m =a(t,L)\,\mu\,h^{\mu-1}$, $g=\Delta f -mh=a(t,L)\,(1-\mu)\, 
h^\mu$, implying
$g \sim a(t,L)^{-1/(\mu-1)} m^{\mu/(\mu-1)}$. Subcases are:\par\noindent
(i)\ $t=0,\ L = \infty$ :  $\Delta f \sim h^{1+1/\delta}$, so $g \sim
m^{1+\delta}$;\par\noindent
(ii)\  $t\ {\rm small},\ L = \infty$, $1 \gg t \gtrsim m^{1/\beta}$:
$\Delta f =a(t)\,h^2, a(t) ={1
\over 2}\chi(t) \sim t^{-\gamma}$, so $g \sim t^\gamma m^2$; 
\par\noindent
(iii)\  $t=0,\ L$ finite, $1 \ll L \lesssim m^{-\nu/\beta}$: $\Delta f
=a(L)\,h^2, a(L) ={1 \over  
2}\chi(L) \sim L^{\gamma/\nu}$, so $g \sim L^{-\gamma/\nu}\,m^2$.   
\begin{figure}
\epsfxsize=8,4cm
\begin{center}
\leavevmode
\epsffile{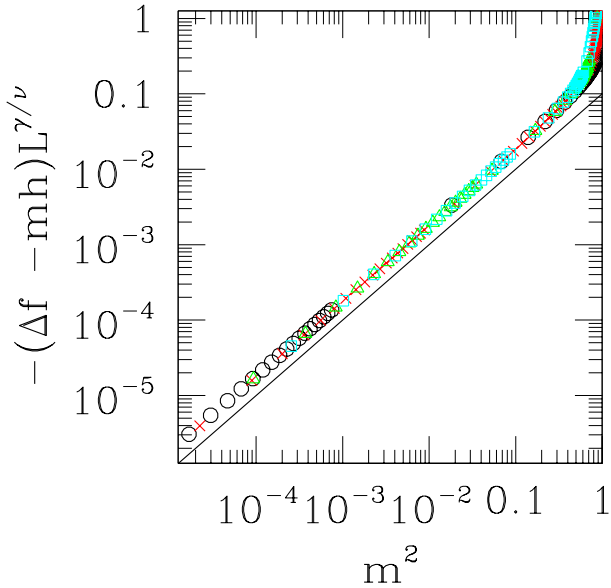}
\caption{
Negative Gibbs free energy $-g= -(\Delta f -mh)$ times
$L^{\gamma/\nu}$ against $m^2$ for pure Ising
systems at criticality. For key to symbols, see caption to 
Fig.~\protect{\ref{fig:fmpure}}. Straight line has unitary slope
and is a guide to the eye.
Normalized magnetizations on horizontal axis only.
}
\label{fig:gpure}
\end{center}
\end{figure}
Case (iii) is depicted in Fig.~\ref{fig:gpure}. One sees that
$-g\,L^{\gamma/\nu} \sim m^2$ as far as $m \simeq 0.6$, which (through
Eq.~(\ref{eq:fhN4})) is consistent with the Gaussian behaviour
predicted for $P_0(M)$ in this case. Close to $m=1$ scaling
breaks down, and the deviation from saturation magnetization must follow
a single-spin-flip picture, $\varepsilon \equiv 1-m \sim \exp(-2/T_c)$.  
The effects of this on $g$ can be worked out from a high-field
expansion, in which ${\cal H}_0$ of Eq.~(\ref{eq:hmag}) is
taken as a perturbation on the field term $h\sum_i
\sigma_i$~\cite{shbe63}. The result is:
\begin{equation}
{dg \over d \varepsilon} = s_0 + {1 \over 2} \ln {1 \over \varepsilon}
+ {\cal O}(\varepsilon)
\label{eq:highfield}
\end{equation}
where $s_0={1 \over 2}\left(\ln 2 - 8/T_c\right)= -1.41617 \dots $ .

\begin{figure}
\epsfxsize=8,4cm
\begin{center}
\leavevmode
\epsffile{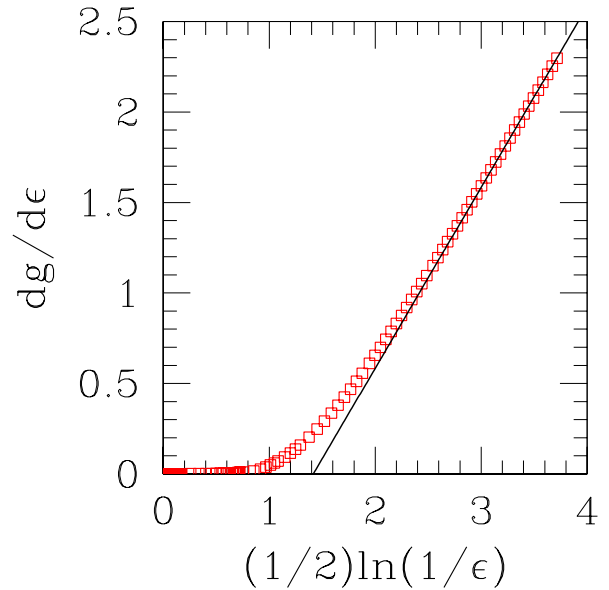}
\caption{
Derivative of Gibbs free energy relative to magnetization deviation
$\varepsilon$ against $(1/2)\ln (1/\varepsilon)$. Points: numerically
calculated derivatives from $L=16$ data for $f$, $m$ ($L=8$
already gives results indistinguishable from those displayed). Straight
line: first two terms on RHS of
Eq.~(\protect{\ref{eq:highfield}}). 
}
\label{fig:hfpure}
\end{center}
\end{figure}

Figure~\ref{fig:hfpure} shows that
Eq.~(\ref{eq:highfield}) is in excellent agreement with numerics already
for $(1/2)\ln (1/\varepsilon) \simeq 2.4\ (\varepsilon \simeq 10^{-2})$.
This provides a rigorous check of our analytic and numerical procedures.
Returning to the connection between $g$ and magnetization distributions,
we first note that, although we are using  $f=L^{-1}\,\ln \Lambda _0$
which holds only for strip length $L_x \to \infty$, the number $N=L_xL$
of spins in Eq.~(\ref{eq:fhN4}) implies a finite,
though possibly very long, strip for the Equation to be of practical use.
An estimate of the error implied by using 
infinite-strip free energies instead of their fully-finite system 
counterparts can be obtained by referring to Table 2 of 
Ref.~\onlinecite{kb93}, where it is shown that for systems with aspect
ratio $\alpha \equiv L_x/L=100$,
the corresponding value of the Binder parameter $Q \equiv \langle M^2
\rangle^2/\langle M^4 \rangle$ is $\simeq 2\%$ off its $\alpha
\to \infty$ limit, $1/3$. As typical widths used
here and, especially in the next section, are in the range $L \lesssim
20$, and assuming that errors in the distribution and in its calculated
moments are of the same order, it follows that using the infinite-strip
expression for $f$ implies deviations in  $P_0(M)$ smaller than  2\% for
$L_x \gtrsim 2,000$. The advantage of this procedure is that $L_x$ can be
seen essentially as a free parameter, i.e. not connected to an actual
number of iterations along the strip. For our purposes here, we shall
always be comparing pure-system results (where $L_x$ is fictitious
in the sense just described) with those obtained on RF strips of the same
width at the same temperature,  
where sampling over disorder typically necessitates an actual $L_x \simeq
10^5$; thus, equating the values of $L_x$ in both systems is both correct
as far as comparisons are concerned and, given the lengths required for
adequate sampling over randomness, fully within acceptable error margins
for the description of pure systems. 

From Eq.~(\ref{eq:fhN4}), for case (iii) where 
$-g\,L^{\gamma/\nu} \sim g_0\,m^2$, $g_0\simeq 10^{-1}$ as shown in
Fig.~\ref{fig:gpure},  
\begin{equation}
P_0(M)=\exp\left[-g_0L_xL^{(\eta-1)}m^2\right]\ \ ,
\label{eq:p0m2}
\end{equation}
where $2-\eta=\gamma/\nu$ was used. Therefore the width of the Gaussian
distribution is ${\cal W} \sim g_0^{-1/2}L^{(1-\eta)/2}/\sqrt{L_x}$.

\subsection{RFIM}
\label{rfim}

We now include the term $\sum_i h_i\sigma_i$ in ${\cal H}_0$ of
Eq.~(\ref{eq:hmag}), with the local fields $h_i$ distributed according to 
Eq.~(\ref{eq:pd}). 

\begin{figure}
\epsfxsize=8,4cm
\begin{center}
\leavevmode
\epsffile{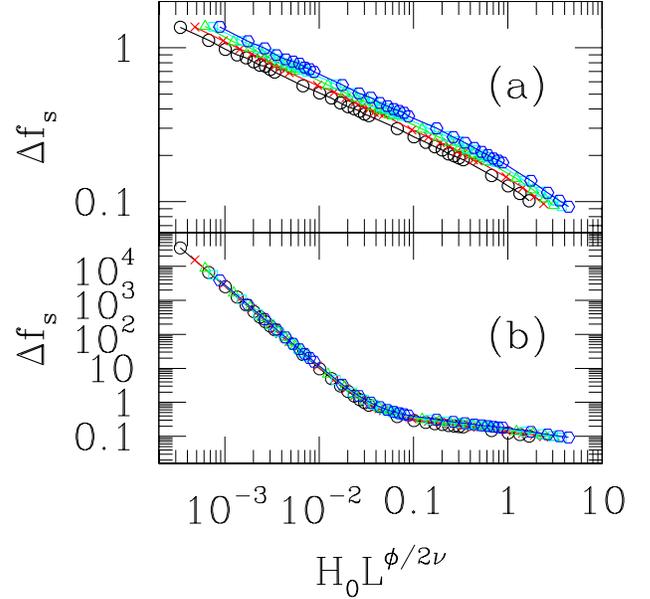}
\caption{$\Delta\,f_s \equiv (\Delta^\prime f(L,H_0) -{\tilde
A}\,L^{-2}\ln
H_0)H_0^{-2d\nu/\phi}$ plotted against $H_0\,L^{\phi/2\nu}$, $\phi=7/4$,
$\nu=1$ .
(a): ${\tilde A}=0$. Bottom to top: $L=4$, $6$, $8$, $10$, and $12$.
(b): ${\tilde A}=10^{-4}$, same notation.     
}
\label{fig:feh0}
\end{center}
\end{figure}

We first consider the application of FSS to the RFIM in zero
uniform field. For bulk systems, theory predicts that the scaling
behaviour of the RFIM depends on $H_0^2 |t|^{-\phi}$ where
$H_0$ is the random-field intensity, $t =(T-T_c(H_0))/T_c(H_0)$ is a
reduced temperature~\cite{fis79,aha83,aha78,fer83,bel85}, and~\cite{aha78}
the crossover exponent is $\phi = \gamma$, the pure Ising susceptibility
exponent. For $d > 2$, $T_c(H_0)$ is the field-dependent temperature at
which a sharp transition still occurs;
in $d=2$ the dominant terms still depend on the same
combination, where now~\cite{bel85} ``$T_c(H_0)$''
denotes a pseudo-critical temperature marking, e.g., the location of
the rounded  specific-heat peak. In $d=2$,
specific heat~\cite{fer83} and neutron-scattering~\cite{bel85} data 
are in good agreement both with the choice of scaling variable as
above, and with the exactly known $\gamma =7/4$ .
For the excess free energy $\Delta^\prime f \equiv f(t,H_0)-f(t,0)$ in two
dimensions, an additive logarithmic
correction arises~\cite{fer83,nvl76}:
\begin{equation}
\Delta^\prime f = A^\ast t^2 \ln H_0 +
H_0^{2d\nu/\phi}\Psi (t\,H_0^{-2/\phi})\ .
\label{eq:bulkrf}
\end{equation} 
In Ref.~\onlinecite{rfcfd}, we showed that the appropriate FSS variable
for the description of correlation functions in finite RF systems at $t=0$
is $x \equiv L\,H_0^{2\nu/\phi}$. While the second term on the RHS of
Eq.~(\ref{eq:bulkrf}) is in that way taken care of, the logarithm needs
separate
consideration. On the basis of renormalization-group arguments, in which
$L^{-1}$ is seen as an additional relevant field~\cite{nig90}, one
realizes that the steps leading to the appearance of the $t^2 \ln H_0$
term in Eq.~(\ref{eq:bulkrf}) also apply here. Indeed, since the
respective  eigenvalue [$y_T=1$ in that case, $y_{L^{-1}}=1$ here]
divides the dimensionality $d=2$~\cite{fer83,nvl76}), a corresponding
scenario obtains at $t=0$ and $L^{-1} \to 0$,
when $L^{-1/\nu}$  is substituted for $t$. Therefore, we assume:
\begin{equation}
\Delta^\prime f(t=0,L,H_0)={\tilde A}\,L^{-2}\ln H_0
+H_0^{2d\nu/\phi}\,{\tilde \Psi}(L\,H_0^{2\nu/\phi})\ .     
\label{eq:frf}
\end{equation}
In Fig.~\ref{fig:feh0}, where $T=T_{c\ 0}$ (thus a small, $H_0$--dependent
shift in ``$T_c(H_0)$''~\cite{fer83,bel85} has been neglected, which
should not matter much for low RF intensities), are displayed results of a
numerical
test of Eq.~(\ref{eq:frf}), both with and without the logarithmic term.

We have found fits of a quality similar to that shown in 
Fig.~\ref{fig:feh0} (b), where ${\tilde A}=10^{-4}$,
for a wide range $10^{-5} \lesssim {\tilde A} \lesssim 10^{-2}$ along
which the $\chi^2$ estimator remains approximately constant. At large 
$H_0\,L^{\phi/2\nu}$, however, the fits deteriorate noticeably (not
obvious
from Fig.~\ref{fig:feh0} (b), because of the large vertical scale), no
doubt owing to the incipient breakdown of the small-RF regime (where, e.g.
the $H_0$--dependent shift in ``$T_c(H_0)$'' is no longer negligible). 
Comparison with experimental data e.g. from Ref.~\onlinecite{fer83} is not
straightforward, as transforming from bulk scaling, Eq.~(\ref{eq:bulkrf}),
to FSS,  Eq.~(\ref{eq:frf}), may involve numerical factors not immediately
available. 

Moving on towards incorporating both RF and uniform field effects, we
again neglect the  $H_0$--dependent shift in ``$T_c(H_0)$'' and make
$T=T_{c\ 0}$. In the presence of several relevant fields $u_1$,$u_2
\dots$ with respective scaling powers  $y_1$,$y_2, \dots$, the
singular part of the free energy scales as~\cite{nvl76}:
\begin{equation}
f(u_1,u_2, \cdots)=
|u_1|^{d/y_1}\,F\left({u_2 \over |u_1|^{y_2/y_1}},{u_3
\over |u_1|^{y_3/y_1}}, \cdots \right)\ . 
\label{eq:fnvl}
\end{equation}
Using $u_1=H_0$, $u_2=L^{-1}$, $u_3=h$, one has in the 
case ($d=2$): $y_1=2\nu/\phi=8/7$, $y_2=1$, $y_3=y_h=15/8$, therefore
\begin{equation}
f(H_0,L,h) = H_0^{2d\nu/\phi}\,F(L\,H_0^{2\nu/\phi}, h\,H_0^{-2 \nu
y_h/\phi})\ .
\label{eq:fLhhrf}
\end{equation}
Possible $\ln H_0$ corrections in the manner of Eq.~(\ref{eq:frf}) have
been omitted, since our interest will focus on the calculation of
the Gibbs free energy, which in the case depends on
$\Delta^{\prime\prime}f = f(H_0,L,h)-f(H_0,L,h=0)$ (see
Eqs.~(\ref{eq:hmag})--(\ref{eq:fhN3})); we are thus
assuming that, at least for small enough $h$, the logarithmic terms cancel
in the subtraction.

Similarly to the pure case but always at $t=0$ and $L^{-1} \to 0$, we
investigate the small--$h$ regime, in which one expects
$\Delta^{\prime\prime}f = a(L,H_0)\,h^\mu$. From Eq.~(\ref{eq:fLhhrf}),
this implies
\begin{equation}
\Delta^{\prime\prime}f = h^\mu\,H_0^{2\nu(d-\mu
y_h)/\phi}\,F_1(L\,H_0^{2\nu/\phi})\ .
\label{eq:frflh}
\end{equation}
By  assuming, as $H_0 \to 0$, a power-law dependence $F_1(x) \sim x^t$,
and demanding that, in this limit, (i) the $H_0$-- dependence of
$\Delta^{\prime\prime}f$ must vanish 
and (ii) the form $h^2\,L^{\gamma/\nu}$ be reobtained, one gets $\mu=2$,
$t=7/4$. Therefore, one has generally for small $h \lesssim L^{-y_h}$:
\begin{equation}
\Delta^{\prime\prime}f = \left({h \over
H_0}\right)^2\,F_1(L\,H_0^{2\nu/\phi})\ .
\label{eq:frf0lh}
\end{equation}
\begin{figure}
\epsfxsize=8,4cm
\begin{center}
\leavevmode
\epsffile{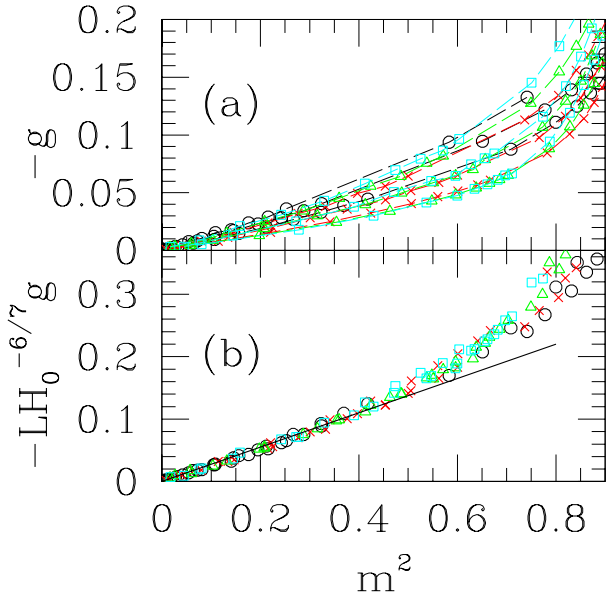}
\caption{
Negative Gibbs free energies: (a) raw data and (b) scaled,  against
squared uniform magnetization. Linear scales used on axes, in order to
underline spread of raw data. In
(a), data for same $(L,H_0)$ are joined
by dashed lines. Strip widths: $L=4$ (circles), $6$ (crosses), $8$
(triangles) and $10$ (squares). RF intensities: $H_0=0.3$, $0.4$, $0.5$
($L=4$); for $L=6$, $8$, $10$,  $H_0=0.2$, $0.3$, $0.4$. For fixed $L$,
$H_0$ increases from bottom to top curves. In (b),
plots of $-L\,H_0^{-6/7}\,g$ collapse well onto a straight line against
$m^2$ (Eq.~(\protect{\ref{eq:ghrflh}}) with $z=1$) up to $m^2 \simeq
0.4$. Straight line in (b) is guide to the eye only, and has slope
$0.275$~.   
}
\label{fig:ghrf}
\end{center}
\end{figure}

Considering now a non-vanishing, but still small, RF, one may assume
crossover to a new power-law form for $F_1(x)$, $F_1(x)  \sim x^z$. 
In this regime ($L\,H_0^{2\nu/\phi} \gtrsim 1$),
\begin{equation}   
\Delta^{\prime\prime}f = L^z\,H_0^{2\nu z/\phi -2}\,h^2\ ,
\label{eq:fhrflh}
\end{equation}   
whence $a(L,H_0)= L^z\,H_0^{2\nu z/\phi -2}$, yielding (see 
Subsection~\ref{pure}):
\begin{equation}
g(L,H_0,m)= -\left[a(L,H_0)\right]^{-1}m^2 = -L^{-z}H_0^{2-2\nu
z/\phi}m^2\ . 
\label{eq:ghrflh}
\end{equation}
Our data for $g(L,H_0,m)$, displayed in Fig.~\ref{fig:ghrf}, are
consistent
with $z=1$, that is, $g(L,H_0,m) \sim -L^{-1}\,H_0^{6/7}\,m^2$~. 

One then has, using Eq.~(\ref{eq:fhN4}), 
\begin{equation}
P_0(M)=\exp\left[-g_1L_xH_0^{6/7}m^2\right]\ \ ,
\label{eq:p0m2rf}
\end{equation}
with $g_1 \simeq 0.28$ from the slope of the straight line in 
Fig.~\ref{fig:ghrf}.
Therefore the distribution is still Gaussian, with a width 
${\cal W} \sim g_1^{-1/2}H_0^{-3/7}/\sqrt{L_x}$. Comparison with a
corresponding pure system, see  Eq.~(\ref{eq:p0m2}) and the arguments
in the paragraph preceding it, gives:
\begin{equation}
{{\cal W}(L,L_x,H_0) \over {\cal W}(L,L_x,0)} \sim
\left(L\,H_0^{8/7}\right)^{-3/8}\ \ \ ,
\label{eq:compw}
\end{equation}  
showing again that the FSS variable $x \equiv L\,H_0^{2\nu/\phi}$ 
is the relevant one. For $x \gg 1$ where RFIM behaviour sets in,
one sees that distribution widths are smaller for RFIM than in zero field.

\section{conclusions}
\label{conc}
We have used TM methods to calculate spin-spin
correlation functions, Helmholtz free energies and
magnetizations on long strips of width $L=3 - 18$ sites of the
two-dimensional RFIM, close to the
zero-field bulk critical temperature.

Through analysis of the probability distributions of correlation functions 
for varying spin-spin distances $R$, we have shown that fits to 
exponential
decay of averaged values against $R$ (for $R$ not too large) give rise to
effective correlation
lengths {\em larger} than in zero field. This is because of the
reinforcement of correlations within domains. At longer distances (i.e.
across many domain walls, $R/L \gg 1$),
fits of exponential decay become unreliable, thus compromising definitions
of effective correlation lengths.   

We have worked out explicit connections between field-- and 
correlation function distributions at high temperatures,
yielding approximate analytical expressions for the
latter. Such expressions account well for trends 
found in numerical data, namely the existence of peaked cusps and the
functional dependence, on $R$ and field intensity $H_0$, of data below the
peak; above the peak, although agreement with numerics is not good, we
have pinpointed that the responsibilty for this lies in a truncation
in our approximate scaling scheme, which decouples scaled
nearest-neighbour interactions from the random field. 
We have discussed the use of analytical expressions, such as the ones
found here, for computation of the corresponding
structure factor. Though results as they stand are far from conclusive, we
have established a rough guide to attempts at connecting basic microscopic
features, such as fluctuations of accumulated fields, to experimentally
observable quantities, e.g. scattering functions.  

We have reanalyzed the
asymptotic behaviour of the relative widths of correlation-function
distributions,
first presented in Ref.~\onlinecite{rfcfd}, and now complemented by
additional data. While our earlier analysis seemed to point
towards exponential saturation,
the new set of data shows that, for fixed $R/L=1$, the
fractional widths of correlation-function distributions behave
consistently with asymptotic power-law saturation, i.e. depending on
$L^{-2.2}$, see Eq.~(\ref{eq:newfit}). The scaling variabls remain as
given previously.

Considering a uniform applied field
$h$, we have derived a connection between Helmholtz free energy $f(h)$,
uniform magnetization $m(h)$, the Gibbs
free energy $g(m)$, and the distribution
function for the uniform magnetization in zero uniform field, $P_0(m)$, 
which is in principle applicable to any finite system. By working at
the bulk zero-field critical temperature $T_{c\ 0}$, we have illustrated
our approach by showing that, for strips, one indeed gets a Gaussian
distribution~\cite{kb93,magdist2} for $m$ not very close 
to saturation. Near $m=1$, where
scaling breaks down and a single-spin-flip picture holds,
a perturbation expansion accounts for the properties of $g(m)$.
Still at  $T_{c\ 0}$, now in  non-zero random field, we have found 
from finite-size scaling and crossover arguments, coupled with numerical
data, that for strip geometries, $P_0(m)$ is still Gaussian, and
its width varies against
(non-vanishing, but small) random-field intensity $H_0$ as
$H_0^{-3/7}$. This is again valid far from saturation (typically,
for $m^2 \lesssim 0.4$, see Fig.~\ref{fig:ghrf}). The ratio between
the width of $P_0(m)$ and the width of the corresponding distribution
for a strip of same length and width in
zero field varies as $(L\,H_0^{8/7})^{-3/8}$.

We expect that at least some of the features discussed here, for
distributions of correlation functions and magnetizations
on strips, translate also for other geometries. Considering, for
instance, square systems:
does the non-trivial form of $P_0(m)$ at bulk criticality in zero 
field~\cite{kb93,magdist1,magdist2} evolve into a corresponding
shape for $H_0 \neq 0$ which depends on the variable
$(L\,H_0^{8/7})^{-3/8}$ as here? 

Finally, recalling possible connections
with experiment: given that the description of correlation decay via
effective correlation lengths runs into difficulties at long distances,
perhaps this compromises naive fits of neutron-scattering data in
the small-wavevector region.

\acknowledgements

R.B.S. acknowledges
partial support from EPSRC Oxford Condensed Matter Theory Rolling Grant
GR/K97783.
S.L.A.d.Q. thanks Department of Theoretical Physics
at Oxford, where this  work was initiated, for the hospitality, and
the cooperation agreement between CNPq and
the Royal Society for funding his visit. 
Research of S.L.A.d.Q.
is partially supported by the Brazilian agencies 
CNPq (grant \# 30.1692/81.5), FAPERJ (grants 
\# E26--171.447/97 and \# E26--151.869/2000) and FUJB-UFRJ.

\end{document}